# Applications of Information Theory to Analysis of Neural Data

Simon R. Schultz[1], Robin A.A. Ince[2] and Stefano Panzeri[2,3]

[1] *Department of Bioengineering, Imperial College London, South Kensington, London SW7 2AZ, UK*

[2] *Institute of Neuroscience and Psychology, 58 Hillhead Street, University of Glasgow, Glasgow G12 8QB, UK*

[3] *Center For Neuroscience and Cognitive Systems, Italian Institute of Technology, Corso Bettini 31 – 38068 Rovereto (Tn) Italy*

8 pages 2638 words.

## Definition

Information theory is a practical and theoretical framework developed for the study of communication over noisy channels. Its probabilistic basis and capacity to relate statistical structure to function make it ideally suited for studying information flow in the nervous system. It has a number of useful properties: it is a general measure sensitive to any relationship, not only linear effects; it has meaningful units which in many cases allow direct comparison between different experiments; and it can be used to study how much information can be gained by observing neural responses in single trials, rather than in averages over multiple trials. A variety of information theoretic quantities are commonly used in neuroscience – (see entry "Definitions of Information-Theoretic Quantities"). In this entry we review some applications of information theory in neuroscience to study encoding of information in both single neurons and neuronal populations.

# Detailed Description

## Information analysis of spike trains to investigate the role of spike times in sensory coding

Mutual information is a widely used tool to study how spike trains encode sensory variables. A typical application of mutual information to spike train analysis is to use it to compare the information content of different representations of neural responses that can be extracted from spike trains. The neural code used by a neuron is often defined operationally as the smallest set of response variables that carries all (or almost all) the information contained in the spike train of the neuron. Mutual information is used to quantify the information content of increasingly complex representations of the neural response, and the simplest representation that carries the most information is chosen as the putative neural code.

An example of this general approach is the investigation of the role of spike times in encoding information. The most established hypothesis on how sensory information is represented in the brain is the spike count coding hypothesis (Adrian, 1928) which suggests that neurons represent information by the number of spikes discharged over some relevant time window. Another hypothesis is the spike timing encoding hypothesis, which suggests that the timing of spikes may add important information to that already carried by spike counts (Rieke et al., 1997; Panzeri et al., 2001). Information theory can be used to understand the role of spike times in carrying sensory information, by using it to characterize the temporal resolution needed to read out the information carried by spike trains. This can be performed by sampling the spike train at different temporal precisions, $\Delta t$, (Fig. 1A) and computing the information parametrically as a function of $\Delta t$ (de Ruyter van Steveninck et al., 1997). The temporal precision required to read the temporal code then can be defined as the largest $\Delta t$ that still provides the full information obtained at higher resolutions. If this precision is equal to the overall length of the window over which neurons carry information, then information is carried only by the number of spikes. As an example, we carried out this type of analysis on the responses of neurons from the VPm thalamic nucleus of rats whose whiskers were stimulated by fast white noise deflections (Montemurro et al., 2007). We found that the temporal precision $\Delta t$ at which neurons transmitted information about whisker deflections was finer than 1 ms (Fig 1B), suggesting that these neurons use high precision spike timing to carry information.

## Information analysis of local field potentials to examine the information content of network oscillations

Information analysis in neuroscience is not limited only to spike train analysis, but it has been used also to study measured of massed population activity, such as Local Field Potentials (LFPs) (Buzsáki et al., 2012). LFPs are operationally defined as the low pass filtered extracellular potential measured by an extracellular intracranial electrode. There are at least three reasons why LFPs are widely used in neuroscience. The first is that they are they are more easily and stably recorded in chronic settings than is the spiking activity of individual neurons. The second is that the LFP captures key integrative synaptic processes and aspects of subthreshold neural activity that cannot be measured by observing the spiking activity of a few neurons alone (Einevoll et al., 2013). The third is that LFPs are more sensitive to network oscillations than measures of spiking activity from small populations. LFPs from a sensory area typically show a power spectrum containing fluctuations over a wide range of frequencies, from <1 Hz to 100 Hz or so. Given that the power of oscillatory activity typically increases during the presentation of a sensory stimulus, many authors have speculated that this oscillatory activity plays a role in brain communication and in particular in sensory-related computations. However, understanding the function of these oscillations has remained elusive and controversial. To gain insights into the function of oscillations in sensory encoding, it is important to understand how they contribute to the representation of the natural sensory environment.

This problem can be addressed by quantifying the oscillation power in any given trial in response to different stimuli, and then computing the information gained by the power at each frequency. Since the power is a continuous variable, the computation of its information is potentially more difficult than the one based on discrete variables like the spike train ones described above. There are at least two ways to solve this problem. The first is to discretize the power in a number of equi-populated bins, and to use bias corrections to eliminate the bias. The second is to fit the data to a parametric distribution. In this case, it is worth reminding that the power computed with most spectral methods follows a chi-square distribution, and thus its square or third can reasonable well approximated by a Gaussian distribution (Magri et al., 2009). This makes the computation of information relatively straightforward. The third potential approach is to use binless methods such as Nearest Neighbors approaches (Kraskov et al., 2004). We tried out these methods extensively on computation of information in power of LFPs, obtaining very similar results with all approaches (see e.g. Magri et al., 2009).

We applied this method to recordings from primary visual cortex of anaesthetised macaques during stimulation with naturalistic colour movies (Belitski et al., 2008; Magri et al., 2012a). This revealed, for the first time, how information about the naturalistic sensory environment is spread over the wide range of frequencies expressed by cortical activity. Although the broad-band nature of the spectrum suggests a contribution to coding from many frequency regions, we found that only two separate frequency regions contribute to coding: the low frequency range and the gamma range (Belitski et al., 2008, see Fig. 2 below). Interestingly, low and high frequency ranges act as perfectly complementary or "orthogonal" information channels: they share neither signal (i.e., stimulus information) nor "noise" (i.e., trial to trial variability for a fixed stimulus). This finding has several implications. First, it shows that, despite the broadband spectrum, only a small number of privileged frequency scales are involved in stimulus coding. Second, it suggests that high-frequency and low-frequency oscillations are generated by different stimulus-processing neural pathways. Third, the finding that different frequency bands code different sensory features in separate, truly independent information channels reinforces the concept of "cortical multiplexing" that we proposed above.

**Information analysis of imaging data to study neural population coding or coupling between different neural signals**

The analysis tools that we have described above have, to date, largely been applied to spike train and time series data recorded using electrophysiological techniques. However, in recent years, imaging technologies have been developed which are capable of resolving neural signaling at systems, cellular and subcellular resolution on a single trial basis (Denk et al., 1990, 1994; Stosiek et al., 2003; Chen et al., 2013). One way to apply information-theoretic tools to the analysis of such imaging data is to convert the data to a "spike train", for instance by applying an algorithm for the detection of action potential evoked calcium transients to calcium imaging data (Oñativia et al., 2013). Such an approach has been used to perform information theoretic analysis of simultaneously recorded populations of cerebellar Purkinje cell complex spikes extracted from *in vivo* calcium imaging movies (Schultz et al., 2009). However, the use of imaging data may also allow a wider set of questions to be approached than can be examined electrophysiologically, by directly examining patterns of pixel intensities.

Another interesting application of information theory to neuroimaging data regards its use for understanding the nature of the coupling between neural activity and fMRI responses. In fact, although there is evidence that fMRI BOLD responses reflect neural activity, it is not clear whether the BOLD signal reflects only the total power of massed neural activity, or only the power in a given band, or rather the relationships between powers of neural activity in different frequency bands. This problem can be cast theoretically into quantifying whether more information about BOLD can be gained from simultaneously observing the power of neural activity in two or more bands of neural activity, than the information gained by observing either band alone. Because mutual information captures all the ways a signal may statistically relate to another, finding that another signal carries extra information demonstrates that this signal truly provides some information that cannot be possibly obtained from the first one. This does not necessarily hold when using methods that capture only specific relationships between signals. For example, an increase in predictability based on linear models may reflect both additional information from the second regressor as well as information that was already present in the first regressor but was not captured by the linear assumption. Application of this idea to simultaneous recording of LFPs and fMRI BOLD in primary visual cortex showed that the beta and alpha band carry information about BOLD that complements that carried by the gamma band, the band that most correlates to the BOLD signal (Magri et al., 2012b).

Since imaging signals such as fMRI have an analogue rather than discrete nature, the practicality of application of information theory to analogue brain signals is crucially dependent upon the development of appropriate regularization and dimensionality reduction algorithms. These might stem from simple yet efficient discretization algorithms (Belitski et al., 2008; Magri et al., 2009), nearest neighbors regularization algorithms (Kraskov et al., 2004), the use of manifold learning techniques for nonlinear dimensionality reduction (Roweis and Saul, 2000; Seung and Lee, 2000; Gan, 2006), and/or the evaluation of information through a decoding step (Quian Quiroga and Panzeri, 2009).

## Acknowledgements

Research supported by the SI-CODE (FET-Open, FP7-284533) project and by the ABC and NETT (People Programme Marie Curie Actions PITN-GA-2011-290011 and PITN-GA-2011-289146) projects of the European Union's Seventh Framework Programme FP7 2007-2013.

## References

Adrian ED (1928) The basis of sensation. New York, NY, USA: Norton & Co. Available at: http://psycnet.apa.org/psycinfo/1928-01753-000 [Accessed January 17, 2014].

Belitski A, Gretton A, Magri C, Marayama Y, Montemurro MA, Logothetis NK, Panzeri S (2008) Low-Frequency Local Field Potentials and Spikes in Primary Visual Cortex Convey Independent Visual Information. J Neurosci 28:5696–5709.

Buzsáki G, Anastassiou CA, Koch C (2012) The origin of extracellular fields and currents — EEG, ECoG, LFP and spikes. Nat Rev Neurosci 13:407–420.

Chen T-W, Wardill TJ, Sun Y, Pulver SR, Renninger SL, Baohan A, Schreiter ER, Kerr RA, Orger MB, Jayaraman V, Looger LL, Svoboda K, Kim DS (2013) Ultrasensitive fluorescent proteins for imaging neuronal activity. Nature 499:295–300.

De Ruyter van Steveninck RR, Lewen GD, Strong SP, Koberle R, Bialek W (1997) Reproducibility and Variability in Neural Spike Trains. Science 275:1805–1808.

Denk W, Delaney KR, Gelperin A, Kleinfeld D, Strowbridge BW, Tank DW, Yuste R (1994) Anatomical and functional imaging of neurons using 2-photon laser scanning microscopy. J Neurosci Methods 54:151–162.

Denk W, Strickler JH, Webb WW (1990) Two-photon laser scanning fluorescence microscopy. Science 248:73–76.

Einevoll GT, Kayser C, Logothetis NK, Panzeri S (2013) Modelling and analysis of local field potentials for studying the function of cortical circuits. Nat Rev Neurosci 14:770–785.

Gan JQ (2006) Feature dimensionality reduction by manifold learning in brain-computer interface design. In: 3rd International Workshop on Brain-Computer Interfaces, Graz, Austria, pp 28–29 Available at: http://cswww.essex.ac.uk/Research/BCIs/BCI06_GAN1.pdf [Accessed January 17, 2014].

Ince RAA, Mazzoni A, Petersen RS, Panzeri S (2010) Open source tools for the information theoretic analysis of neural data. Front Neurosci 4:62–70.

Kraskov A, St\ögbauer H, Grassberger P (2004) Estimating mutual information. Phys Rev E 69:66138.

Magri C, Mazzoni A, Logothetis NK, Panzeri S (2012a) Optimal band separation of extracellular field potentials. J Neurosci Methods 210:66–78.

Magri C, Schridde U, Murayama Y, Panzeri S, Logothetis NK (2012b) The Amplitude and Timing of the BOLD Signal Reflects the Relationship between Local Field Potential Power at Different Frequencies. J Neurosci 32:1395–1407.

Magri C, Whittingstall K, Singh V, Logothetis NK, Panzeri S (2009) A toolbox for the fast information analysis of multiple-site LFP, EEG and spike train recordings. BMC Neurosci 10:81.

Montemurro MA, Panzeri S, Maravall M, Alenda A, Bale MR, Brambilla M, Petersen RS (2007) Role of Precise Spike Timing in Coding of Dynamic Vibrissa Stimuli in Somatosensory Thalamus. J Neurophysiol 98:1871–1882.

Oñativia J, Schultz SR, Dragotti PL (2013) A finite rate of innovation algorithm for fast and accurate spike detection from two-photon calcium imaging. J Neural Eng 10:046017.

Panzeri S, Petersen RS, Schultz SR, Lebedev M, Diamond ME (2001) The Role of Spike Timing in the Coding of Stimulus Location in Rat Somatosensory Cortex. Neuron 29:769–777.

Quian Quiroga R, Panzeri S (2009) Extracting information from neuronal populations: information theory and decoding approaches. Nat Rev Neurosci 10:173–185.

Rieke F, Bialek W, Warland D, de Ruyter van Steveninck RR (1997) Spikes: Exploring the Neural Code. Bradford Book.

Roweis ST, Saul LK (2000) Nonlinear Dimensionality Reduction by Locally Linear Embedding. Science 290:2323–2326.

Schultz SR, Kitamura K, Post-Uiterweer A, Krupic J, Häusser M (2009) Spatial Pattern Coding of Sensory Information by Climbing Fiber-Evoked Calcium Signals in Networks of Neighboring Cerebellar Purkinje Cells. J Neurosci 29:8005–8015.

Seung HS, Lee DD (2000) The Manifold Ways of Perception. Science 290:2268–2269.

Stosiek C, Garaschuk O, Holthoff K, Konnerth A (2003) In vivo two-photon calcium imaging of neuronal networks. Proc Natl Acad Sci 100:7319–7324.

# Figures and Figure Captions

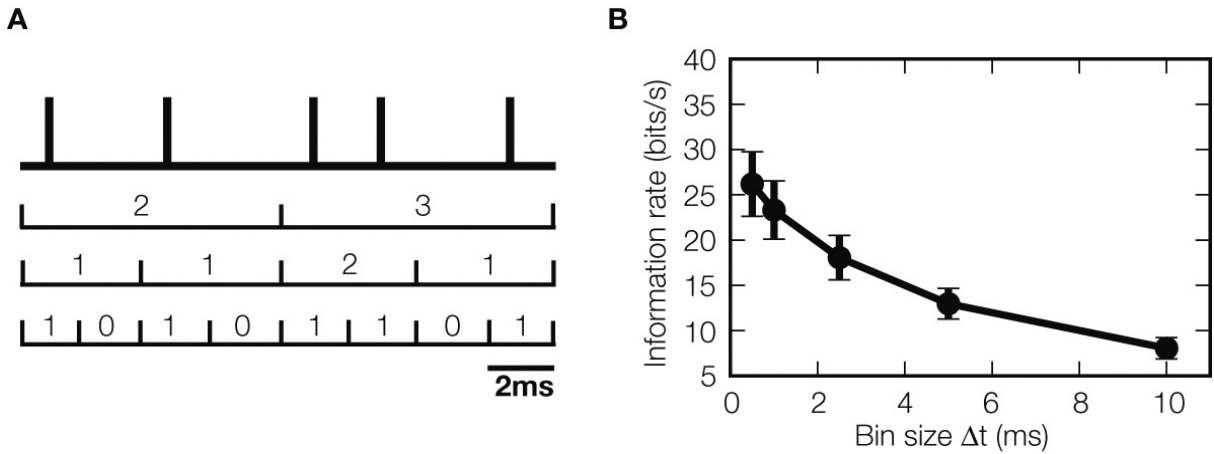

**Figure 1**: *Effect of temporal resolution of spike times on information*. (A): The response of a neuron is initially recorded as a series of spike times. To investigate the temporal resolution at which spike times carry information, the spike train is binned at a variety of different time resolutions, by labeling the response at each time with the number of spikes occurring within that bin, thereby transforming the response into a discrete integer sequence. (B): The information rate (information per unit time) about whisker deflections carried by VPm thalamic neurons as a function of bin width, Δt, used to bin neural responses. Information rate increased with bin resolution up to 0.5ms, the limit of the experimental setup. This shows that a very fine temporal resolution is needed to read out the sensory messages carried by these thalamic spike trains. Figure reprinted with permission from (Ince et al., 2010).

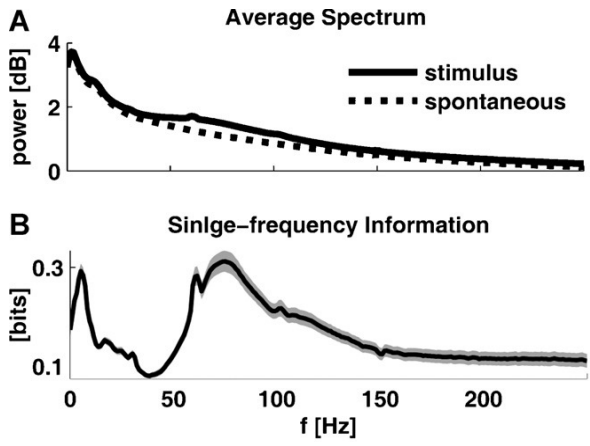

**Fig. 2**: *The visual information carried by LFP power at different frequencies.* A) LFP power spectrum of V1 recordings anaesthetized macaques during either spontaneous activity in the dark (dashed line) or during the presentation of a color movie stimulus (solid line). (B) Information about the movie stimulus carried by LFP power at different frequencies. The area indicates the SEM. Reproduced from (Magri et al., 2012a).